\documentclass{jot}

\usepackage{multirow}
\usepackage{balance}

\usepackage{caption,subcaption}
\usepackage{textcomp}
\usepackage{float}
\usepackage{multicol}

\usepackage{listings}
\usepackage{color,xcolor}
\usepackage{tikz}
\newcommand*\circled[1]{\tikz[baseline=(char.base)]{
    \node[shape=circle,draw,inner sep=1pt] (char) {#1};}}

\usepackage{pgfplots}
\pgfplotsset{compat=newest}
\usetikzlibrary{patterns}

\newcommand{\jit}{JIT}
\newcommand{\vm}{VM}

\newcommand{\astree}{AST}
\newcommand{\rpython}{RPython}
\newcommand{\truffle}{Truffle/Graal}

\newcommand{\hotspot}{HotSpot\textsuperscript{\texttrademark}}
\newcommand{\lif}{language implementation framework}
\newcommand{\Lif}{Language implementation framework}

\newcommand{\tstack}{\texttt{traverse\_stack}}
\newcommand{\comment}[1]{}
\newcommand{\memo}[1]{}

\jotdetails{
    volume=vv,      
    number=nn,       
    articleno=aa,   
    year=yyyy,      
    license=ccbyncnd    
  }

\articletype{regular}

\title{Threaded Code Generation with a Meta-Tracing JIT Compiler}

\author[$\ast$]{Yusuke Izawa}
\author[$\ast$]{Hidehiko Masuhara}
\author[$\dagger$]{Carl Friedrich Bolz-Tereick}
\author[$\ast$]{Youyou Cong}

\affil[$\ast$]{Tokyo Institute of Technology, Japan}
\affil[$\dagger$]{Heinrich-Heine-Universität Düsseldorf, Germany}

\runningauthor{Izawa \textit{et al.}}


\runningtitle{Threaded Code Generation with a Meta-Tracing JIT Compiler} 

\runningauthor{Izawa \textit{et al.}}

\begin{abstract}
  Language implementation frameworks, e.g., RPython and Truffle/Graal, are practical tools
for creating efficient virtual machines, including a well-functioning just-in-time (JIT)
compiler. It is demanding to support multitier JIT compilation in such a framework
for language developers.  This paper presents an idea to generate threaded code by
reusing an existing meta-tracing JIT compiler, as well as an interpreter design for
it. Our approach does not largely modify RPython itself but constructs an effective
interpreter definition to enable threaded code generation in RPython. We expect our system
to be extended to support multilevel JIT compilation in the RPython framework. We
measured the potential performance of our threaded code generation by simulating its
behavior in PyPy. We confirmed that our approach reduced code sizes by 80 \% and
compilation times by 60 \% compared to PyPy's JIT compiler on average, and ran about 7 \%
faster than the interpreter-only execution.
\end{abstract}

\acknowledgment{We would like to thank the reviewers of the ICOOOLPS 2021 workshop
  for their valuable comments. This work was supported by JSPS KAKENHI grant number
  18H03219, 21J10682, and JST ACT-X grant number JPMJAX2003.}

\begin{document}
\maketitle
\urlstyle{rm}

\section{Introduction}


\Lif{}s, such as \rpython{}~\cite{Bolz2009} and \truffle{}~\cite{Wurthinger2012}, help
language developers in building a full-fledged
virtual machine with a smaller amount of implementation effort.  Those \lif{}s have
a mechanism that takes an interpreter definition of a language
and yields a virtual machine (\vm{}) with advanced features, including a quality
just-in-time (\jit{}) compiler. The effectiveness and usefulness of \lif{}s are
demonstrated by efficient implementations of many programming language implementations
including PyPy~\cite{Bolz2009}, GraalPython~\cite{graalpython}, Topaz~\cite{topaz},
TruffleRuby~\cite{truffleruby}, RSqueak~\cite{Felgentreff:2016:BHV:2991041.2991062}, and
TruffleSqueak~\cite{Niephaus:2019:GTS:3357390.3361024}. Not only those implementations are
realized by writing interpreters, many of them exhibit better performance than their
interpreter-based counterparts (e.g., CPython and CRuby).

It is challenging for \lif{}s to support multitier \jit{} compilation.
Multitier \jit{} compilation is a technique that compiles different
parts of programs at different optimization levels to balance
between compilation overheads and the efficiency of compiled
code\footnote{For example, the Jalape\~{n}o Java \vm{} has a baseline
  and an optimizing compiler with three optimization
  levels~\cite{Alpern:10.1145/320384.320418}. In addition, The four tire JIT in the JavaScript
  engine in Webkit~\cite{javascriptcore} has four different optimization levels.}.
Na\"{i}vely, multitier
compilation requires a compiler for each optimization level.  It is also possible
to construct one compiler that can yield code at different optimization levels, but
implementing such a compiler would require more effort of language developers.
Since \lif{}s need to \emph{generate} such a compiler from interpreter definitions, it is
more challenging to support multitier compilation with the same interpreter definition.

In this paper, we propose a technique that generates threaded code by
reusing an existing meta-tracing \jit{} compiler, namely \rpython{}.
Threaded code generation is a compilation technique that simply
converts each operation of a source program into a call to a
respective handler function. Some multitier \vm{}s use threaded code
generation as the baseline compiler since its compilation speed is
extremely fast. Our idea is to use existing \rpython{}'s engine
(i.e., the meta-tracing compilation machinery) for threaded code
generation so that it will serve as the baseline \jit{} compiler for
\rpython{}-based language implementations.
We expect that our threaded code generation should be placed between an interpreter
execution and a tracing JIT execution since it is just baseline compilation. Given that
context, the threaded code compilation should reduce a compilation code size to compile
it fast.
Although this paper focuses on threaded code generation, we hope that our approach would
be able to be extended to compilation at different optimization levels in the future.

The proposal of the paper is positioned as an application of our
\textit{meta-hybrid \jit{} compiler framework}
project~\cite{Izawa:2020:10.1145/3426422.3426977} to a production-level \lif{}, namely the
RPython-backend for PyPy~\cite{Rigo2006:10.1145/1176617.1176753} in the context of
threaded code generation. In contrast to our original
proposal~\cite{Izawa:2020:10.1145/3426422.3426977} that is based on a simple experimental
\lif{}, this paper realizes threaded code generation on a production level \lif{} by
mostly reusing the existing implementation that does not consider threaded code generation
or other levels of optimization at all. In other words, this approach doesn't need to
modify the \rpython{}'s compilation engine too much, but we realize it just by preparing a
specific interpreter definition and implement a new trace compilation engine that share
almost all the code base (details are explained in
Section~\ref{sec:compilation-tactic-of-baseline-jit}).

In this paper, we make the following contributions:

\begin{itemize}
\item an idea to build a method-based threaded code generator on
  top of \rpython{},
\item an implementation design to realize a method-based threaded code generation
  with a meta-tracing \jit{} compiler, and
\item measuring the potential performance of the generated threaded code through
  preliminary experiments in PyPy.
\end{itemize}

The rest of this paper is organized as follows. Section~\ref{sec:background} shows the
background. Section~\ref{sec:compilation-tactic-of-baseline-jit} explains the idea of
realizing a method-based baseline JIT compiler on top of \rpython{}.
In Section~\ref{sec:discussion}, through preliminary benchmark experiments, we discuss
how well our threaded code generation performs in practice, and what kind of programs it
should be applied to. Section~\ref{sec:relatedwork} presents the related work and
Section~\ref{sec:conclusion} concludes this paper.

\section{Background}
\label{sec:background}

This section briefly gives an overview of a \lif{}, and the meta-tracing \jit{} compiler
in PyPy/RPython as well as threaded code.

\subsection{Language Implementation Framework}
\label{sec:lang_impl_framework}

A \lif{} is a tool that generates a high-performance \vm{} from an interpreter
definition. In a traditional development way, programming language developers have to
implement \vm{} components, such as an interpreter, \jit{} compiler, memory management
model, etc., from scratch for each language. However, by using a \lif{} developers need to
write only an interpreter by using a \lif{} when they build a language.

There are two state-of-the-art frameworks called \rpython{} and
\truffle{}. \rpython{}~\cite{Rigo2006:10.1145/1176617.1176753} is a part of the
PyPy project; PyPy is generated from the \rpython{} framework. On the other hand,
\truffle{}~\cite{Wurthinger2012} is a part of GraalVM project that is being developed by
Oracle Lab. They are successful in generating high-performance language implementations
for Python~\cite{PyPywebsite, graalpython}, PHP~\cite{hippyvm}, Ruby~\cite{topaz,
  truffleruby}, R~\cite{fastr}, and so on.

\rpython{} and \truffle{} require different interpreter definition styles; a bytecode
interpreter and an abstract-syntax-tree (\astree{}) interpreter, respectively. In
addition, to enable \jit{} compilation and other optimizations, framework users have to
follow the implementation manners that are provided by the frameworks. For example, at least,
\rpython{} users have to write hint functions at a right place (details are shown in
Section~\ref{sec:meta_tracing_jit}), and \truffle{} users should define \astree{} nodes by
inheriting \texttt{Node} class which provides \truffle{} and override \texttt{execute} method
inside their defined \astree{} nodes.

\subsection{PyPy/RPython and Meta-tracing JIT Compiler}
\label{sec:meta_tracing_jit}

PyPy is an implementation of Python language, based on the
RPython~\cite{Rigo2006:10.1145/1176617.1176753} compiler. It has a high-performance
tracing JIT compiler, which is not directly implemented but generated by the RPython
compiler.  The RPython compiler accepts a bytecode interpreter written in RPython. A
meta-tracing JIT compiler~\cite{Bolz2009} keeps track of the execution of a user-defined
interpreter and compiles a hot loop of a target language.

\begin{lstlisting}[language=Python,caption={A simple example of a bytecode interpreter
written in RPython},label={lst:rpython-interp},float=t]
jitdriver = JitDriver(reds=['self'],
                greens=['pc', 'bytecode'])

def interp(self):
    pc = 0
    while True:
        jitdriver.jit_merge_point(
            self=self,pc=pc,
            bytecode=bytecode)
        opcode = bytecode[pc]
        pc += 1
        if opcode == ADD:
            ...
        elif opcode == JUMP:
            t = ord(bytecode[pc])
            pc += 1
            if t < pc:
                jitdriver.can_enter_jit(
                    self=self,pc=t,
                    bytecode=bytecode)
            pc = t
        ...
\end{lstlisting}

Listing~\ref{lst:rpython-interp} is an example definition that a language developer needs
to write in RPython. Note that a language developer needs to define
\texttt{jitdriver} for telling the necessary information to RPython's meta-tracing \jit{}
compiler. There are also special functions, namely \texttt{jit\_merge\_point} and
\texttt{can\_enter\_jit}. \texttt{jit\_merge\_point} and \texttt{can\_enter\_jit} should
be placed at the beginnings of a bytecode dispatch loop and where back-edge instruction
occurs (e.g., the end of \texttt{JUMP} definition in Listing~\ref{lst:rpython-interp}),
respectively.

\subsection{Threaded Code}
\label{sec:threaded-code}

\begin{figure}[t!]
  \centering
  \includegraphics[width=.8\linewidth]{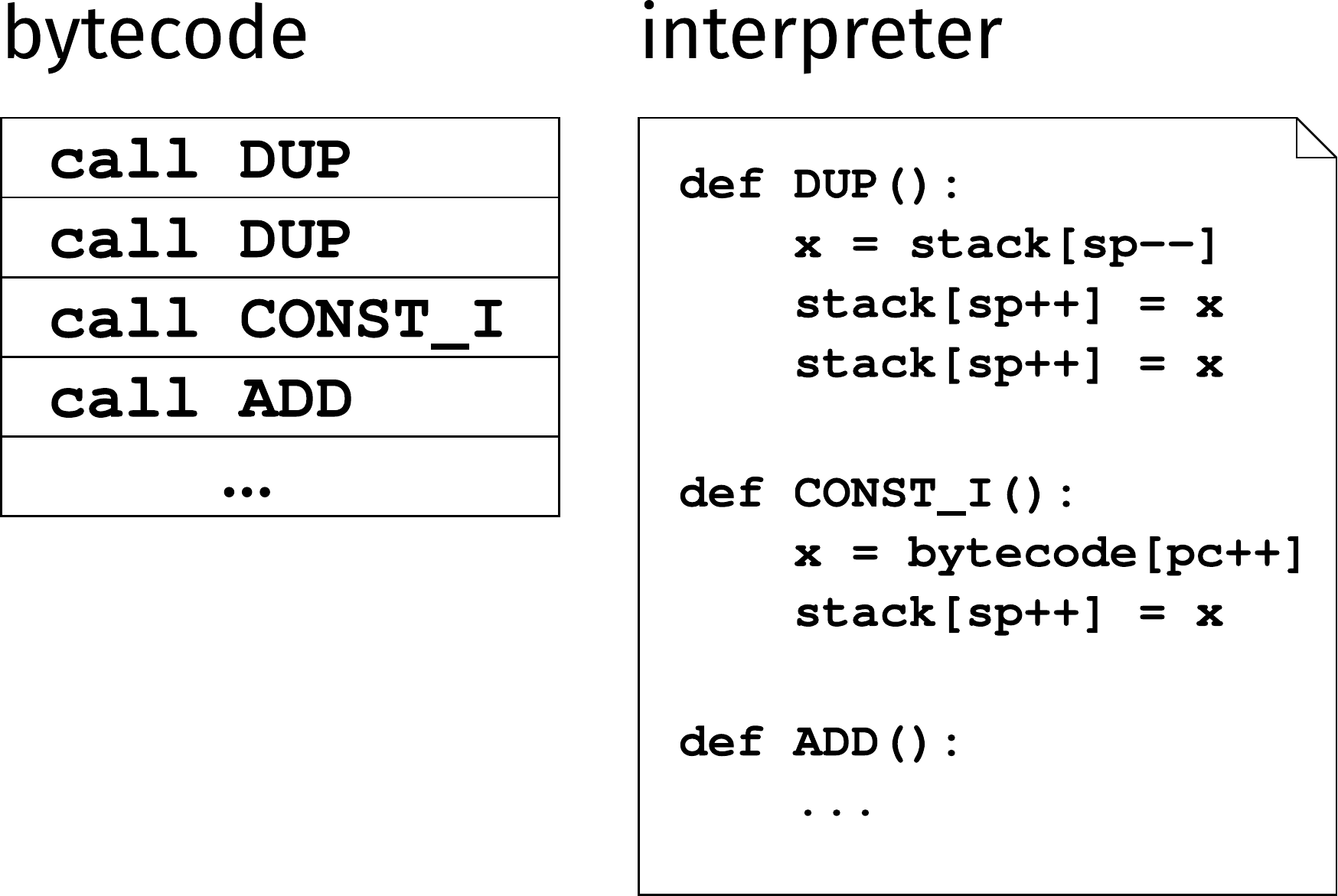}
  \caption{An overview of how threaded code works.}
  \label{fig:overview-threaded-code}
\end{figure}

Threaded code~\cite{Bell:1973:10.1145/362248.362270, Hong:10.1145/146559.146561} is a
technique to improve the performance of a bytecode interpreter. The interpreter separately
defines \emph{handler functions} for all bytecode instructions as shown on the right-hand
side in Figure~\ref{fig:overview-threaded-code}. A program is a sequence of call
instructions to handlers as shown on the left-hand side. Executing a threaded code-based
program reduces the number of indirect branching that significantly pose a performance
penalty at runtime because of branch mispredictions~\cite{ertl2003the}.


\begin{figure*}[t!]
  \centering
  \includegraphics[width=\linewidth]{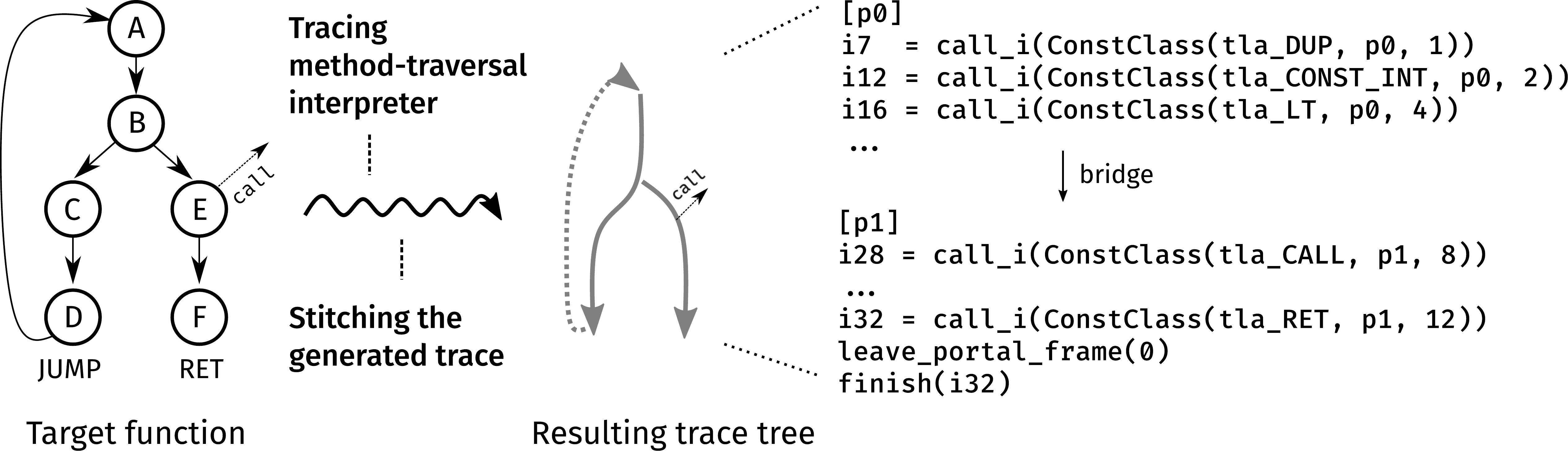}
  \caption{A sketch of how \rpython{} method-based baseline \jit{} compiler
    works. From the target function in the left-hand side, it generates the trace
    tree shown in the right-hand side.}
  \label{fig:overview_rpython_baseline}
\end{figure*}

\section{The Compilation Tactic of RPython's Baseline JIT}
\label{sec:compilation-tactic-of-baseline-jit}

In this section, we present how to realize method-based baseline \jit{} strategy
on top of \rpython{} without implementing a compilation engine from scratch.

The objective of introducing threaded code is for less start-up and compilation time in
the \rpython{}\footnote{An alternative approach to improve warm-up performance is to
  improve the dispatching mechanism of an interpreter, for example, by using threaded
  jumps. It would not be easy to realize such approaches in RPython as it currently
  assumes more straightforwardly written interpreters.}. In general, a tracing JIT
compiler automatically inlines function calls
and applies several optimizations to a trace. The longer the trace you get and the better
native code you want to generate, the longer the compilation time. In contrast, threaded
code generation only leaves the call instruction to a subroutine, so tracing doesn't
consume much time. We also apply only simple optimizations such as constant-folding and
removing duplicated operations to the obtained trace from the
threaded code generator, so we can reduce the compilation time than a normal tracing mode.

\subsection{The Compilation Principle}
\label{sec:comilation-principle}

Our threaded code generation is achieved by carefully controlling the \rpython{}'s
meta-tracing compiler and reconstructing a control flow from the resulted trace. We
realize it just by preparing a model of a specific interpreter definition (called
\emph{method-traversal interpreter}) and a new trace compilation mechanism (called
\emph{trace-stitching}). Below, we explain our approach by comparing our approach against
a typical \jit{} compilation process in \rpython{}.

When RPython compiles a base-program executed by an interpreter, it
\begin{description}
\item[starts compiling] at the beginning of a
  loop, which is dynamically detected;
\item[for each operation in the base program,] follows into the
  respective handler body in the interpreter, which effectively
  eliminates ``interpretation'' (i.e., code dispatching and operand
  manipulation) by the interpreter;
\item[at a function call in the base program,] follows into the body of
  the callee function, which effectively achieves function inlining;
\item[at a conditional expression in the base program,] follows only
  one of the branch with emitting a \emph{guard} operation for other
  branches; and
\item[at a conditional branch in the handler] (including selection
  of an arithmetic operation based on operands' runtime types),
  traces only one of the branch to achieve effective type-specialization;
\item[finishes compiling] at the end of the loop.
\end{description}

Our threaded code generation operates the \rpython{} compiler so that it
\begin{description}
\item[starts compiling] at the beginning of a method/function in the
  base-program\footnote{Whether the system uses threaded code generation
    or not is an open issue that we will consider in the future.  For the time
    being, we merely assume that the threaded code generator is
    invoked for a particular base-program method/function.};
\item[for each operation in the base-program,] follows the code
  dispatching part of the interpreter, but does not trace into the
  handler body but emits a call instruction to the respective handler;
\item[at a function call in the base-program,] emits a call instruction
  and continues tracing of the operations after the functional call;
\item[at a conditional expression in the base-program,] follows \emph{all}
  branches;
\item[at a conditional branch in the handler] --- this will not
  happen since the compiler does not trace the inside of handlers;  and
\item[finishes compiling] at the end of the method/function.
\end{description}

To drive the RPython compiler like that, our proposal consists of the following
three techniques:

\begin{description}
\item [The \emph{method-traversal interpreter} technique.] We write an
  interpreter to let the tracing mechanism of RPython traverse
  all execution paths in a base-program method/function.  We achieve this behavior
  by merely defining the interpreter in a specific way, but
  not modifying the existing RPython infrastructure.

\item [The hinting technique.] We let the \rpython{} compiler not trace inside of
  handlers and the callee of a function/method call in a base-program. This is also
  achieved by placing existing RPython annotations into the interpreter definition.

\item [The \emph{trace stitching} technique.] We reconstruct the original control
  flow of a base-program function/method from a recorded trace.  Since
  the method-traversal interpreter technique will yield a straight-line trace that covers
  all the execution paths, this technique will split the trace into basic blocks and then
  connect them together by using branch and jump instructions.  This is achieved by adding
  a post-processing module into the \rpython{} tracer.
\end{description}



\begin{figure}[t!]
  \centering
  \includegraphics[width=.4\linewidth]{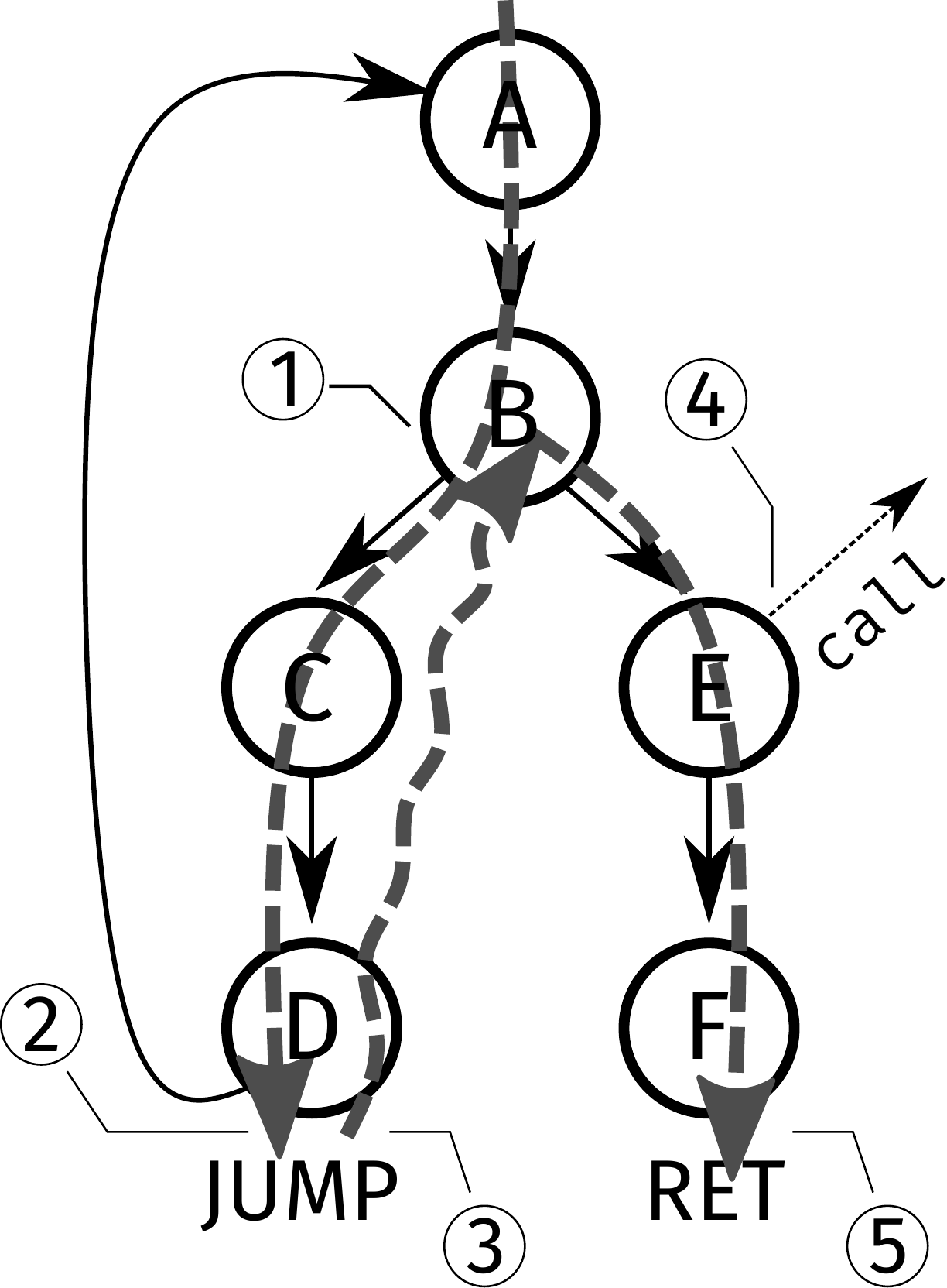}
  \caption{Tracing the entire of a function with method-traversal interpreter.}
  \label{fig:tracing_method_traversal_interp}
\end{figure}

\begin{lstlisting}[
language=Python,float=t,
caption=Skeleton of method-traversal interpreter and subroutines decorated with
    \texttt{dont\_look\_inside}.,
label={lst:skeleton_method_traversal_interp}]
@dont_look_inside
def tla_ADD(self, pc):
    x, y = self.pop(), self.pop()
    self.push(y.add(x))
    return pc

@dont_look_inside
def tla_CONST_INT(self, pc):
    arg = ord(self.bytecode[pc])
    self.push(W_IntObject(int(arg)))
    return pc + 1

driver = JitDriver(reds=['self'],
              greens=['pc','bytecode','traverse_stack'])

class Frame:
    def interp(self, pc, traverse_stack):
        while True:
            driver.jit_merge_point(
                bytecode=self.bytecode,pc=pc,self=self,
                traverse_stack=traverse_stack)
            opcode = ord(self.bytecode[pc])
            pc += 1
            if opcode == ADD:
                pc = self.tla_ADD(pc)
            elif opcode == JUMP:
                ...
            elif opcode == RET:
                ...
            elif opcode == JUMP_IF:
                ...
\end{lstlisting}

Figure~\ref{fig:overview_rpython_baseline} shows a high-level example of \rpython{}
baseline \jit{} compiler. The left-hand side of
Figure~\ref{fig:overview_rpython_baseline} represents the control
flow of a target function. B -- C -- E is a conditional branch, D is a back-edge
instruction, and F is a return. The compiler finally generates a trace
tree~\footnote{Each trace has a linear control flow, but they are compiled as a
  bridge.}, which covers a function body as shown in the right-hand side of
Figure~\ref{fig:overview_rpython_baseline}. In contrast to trace-based compilation,
it keeps the original control flow, we can see that the bodies of subroutines are not
inlined but call instructions to them are left.

To produce such a trace tree, the tracer of \rpython{} baseline \jit{} has to sew and
stitch generated traces. We call this behavior \textit{trace tailoring}. Technically
speaking, the compiler traces a special instrumented interpreter namely
\textit{method-traversal interpreter}. Since the obtained trace from the
method-traversal interpreter ignores the original control flow, we have to restore
it. To rebuild the original control flow, in the next phase, the baseline \jit{}
compiler stitches the generated trace. We call this technique \textit{trace
  stitching}. In the next sections, we will explain method-traversal interpreter and
trace stitching, respectively.


\begin{figure}[t]
\noindent\begin{minipage}[t]{.48\linewidth}
\begin{lstlisting}[language=Python, caption={An example bytecode with the
control flow shown in Figure~\ref{fig:tracing_method_traversal_interp}.},
label={lst:method-traversal-example-bytecode},]
DUP,
CONST_INT, 1,
GT,
JUMP_IF, 10,
CONST_INT, 1
SUB,
JUMP, 0
CALL, 23,
EXIT,
\end{lstlisting}
\end{minipage}\hfill
\begin{minipage}[t]{.48\linewidth}
\begin{lstlisting}[language=Python, caption={An example program corresponding to
Listing~\ref{lst:method-traversal-example-bytecode}.},label={lst:method-traversal-ex-program}]
while True:
    if x > 1:
        x -= 1
    else:
        x = call g(x)
        return x
\end{lstlisting}
\end{minipage}
\end{figure}

\subsection{Method-traversal Interpreter}\label{sec:method_traversal_interp}

\begin{lstlisting}[language=Python,caption={Definition of
JUMP\_IF.},label={lst:def_jump_if},float=t]
if opcode == JUMP_IF:
     target = ord(self.bytecode[pc])
     e = self.pop()
     if self._is_true(e):
         if we_are_jitted():
             pc += 1
             # save another direction
             traverse_stack = t_push(
                 pc, traverse_stack)
         else:
             if t < pc:
                driver.can_enter_jit(pc=target,
                  bytecode=self.bytecode,self=self,
                  traverse_stack=traverse_stack)
         pc = target
     else:
         if we_are_jitted():
             # save another direction
             traverse_stack = t_push(target,
                 traverse_stack)
         pc += 1
\end{lstlisting}

\begin{lstlisting}[language=Python,caption={Definition of JUMP.},label={lst:def_jump},
float=t]
@dont_look_inside
def cut_here(self, pc):
    "A pseudo function for trace stitching"
    return pc

if opcode == JUMP:
    t = ord(self.bytecode[pc])
    if we_are_jitted():
        if t_is_empty(traverse_stack):
            pc = t
        else:
            pc, traverse_stack = traverse_stack.t_pop()
            # call pseudo function
            pc = cut_here(pc)
    else:
       if t < pc:
           jitdriver.can_enter_jit(
               bytecode=self.bytecode,pc=t,self=self,
               traverse_stack=traverse_stack)
       pc = t
\end{lstlisting}

\begin{lstlisting}[language=Python,caption={Definition of RET.},label={lst:def_ret},
float=t]
if opcode == RET:
    if we_are_jitted():
        if t_is_empty(traverse_stack):
            return self.tla_RET(pc)
        else:
            pc, traverse_stack = traverse_stack.t_pop()
    else:
        return self.tla_RET(pc)
\end{lstlisting}

\begin{figure*}[t]
  \centering
  \includegraphics[width=\linewidth]{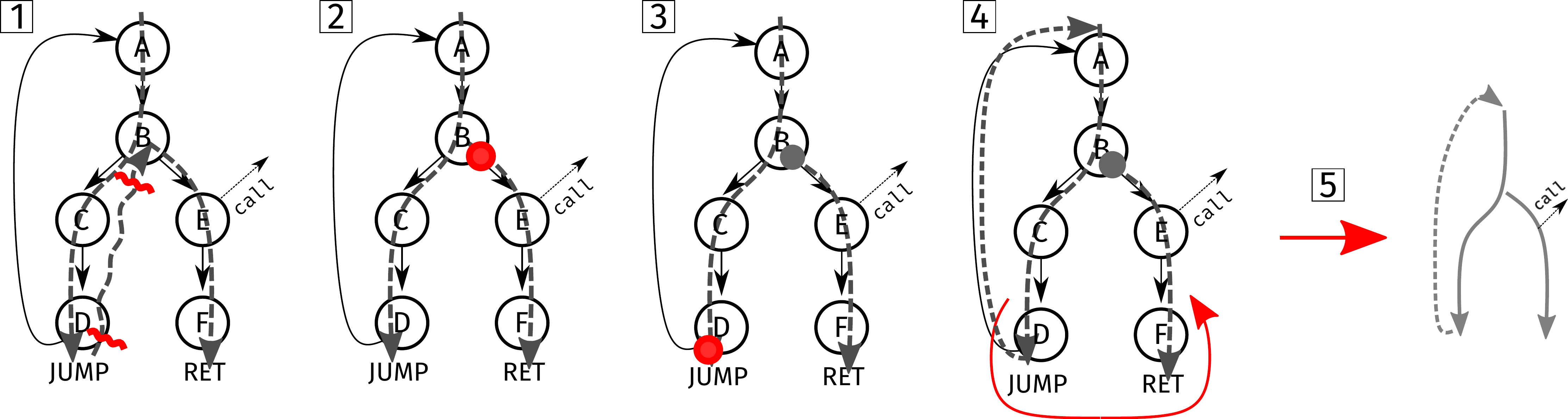}
  \caption{The working flow of trace stitching.}
  \label{fig:overview_trace_stitching}
\end{figure*}

We propose method-traversal interpreter, a specially instrumented interpreter for the
baseline \jit{} compiler. It works as an abstract interpreter because it follows complete
control flow graph by exploring both sides of a conditional branch.

The skeleton of method-traversal interpreter is shown in
Listing~\ref{lst:skeleton_method_traversal_interp}. All handlers that are shown at the top
of the listing, are decorated by \texttt{dont\_look\_inside} hint which tells the tracer
not to trace the function body. Furthermore, specific areas are written in the \texttt{then}
block of \texttt{we\_are\_jitted}. This hint function returns \texttt{True} after entering
tracing. Therefore, the resulting trace has only call instructions to subroutines.

Figure~\ref{fig:tracing_method_traversal_interp} shows how method-traversal interpreter
traverses a function body with respect to the bytecode denotes in
Listing~\ref{lst:method-traversal-example-bytecode}
and~\ref{lst:method-traversal-ex-program}. In
Figure~\ref{fig:tracing_method_traversal_interp}, the gray-colored dotted line means a
generated trace with the method-traversal interpreter. Normally, a tracing \jit{} only
follows an executed side of the conditional branch. In contrast, the baseline \jit{} tracer
follows the both sides. To enable it, method-traversal interpreter manages a special
stack data structure called \texttt{traverse\_stack}. It only stores program counters, so
it is marked as \textit{green} and finally removed from the resulting trace.

We explain the behavior of method-traversal interpreter with respect to the
examples. The differences from a normal tracing \jit{} compiler are: (1) conditional
branch, (2) back-edge instruction, (3) function call, and (4) function return.

\subsubsection{Conditional branch.}

Our baseline \jit{} tracer follows both sides of a conditional branch; firstly,
tracing \texttt{then} branch, and tracing \texttt{else} branch next.

\paragraph*{When tracing a conditional branch \circled{1} in
  Figure~\ref{fig:tracing_method_traversal_interp},} it saves the program counter in
another direction of a conditional branch to the
\texttt{traverse\_stack}. Listing~\ref{lst:def_jump_if} shows the handler for the
\texttt{JUMP\_IF}. You can see that \tstack{} saves another directions in lines 8 and 19.

\subsubsection{Back-edge instruction.}

Upon a back-edge instruction, the baseline \jit{} tracer jumps to one of the remaining branches.
\rpython{}'s original tracer follows a back-edge instruction and finishes tracing when it
reaches the beginning of tracing. We modify such a behavior not to finish tracing until
the tracer reaches the end of a target method and visits its all paths.

\paragraph*{When tracing a back-edge instruction at \circled{2},} it does not follow the
jump target. Instead, at \circled{3}, it pops a program counter from \tstack{} and goes to
the other branch which is an unfollowed branch of a previous conditional jump (E in the
Figure~\ref{fig:overview_rpython_baseline}).

Seeing the implementation of \texttt{JUMP} in Listing~\ref{lst:def_jump}, before
jumping to somewhere, it checks whether \tstack{} is empty or not. If empty, the baseline
tracer normally executes \texttt{JUMP}. Otherwise, it restores the saved program counter
from \tstack{} and goes to that place. To tell the place of a back-edge instruction,
we have to call a pseudo function \texttt{cut\_here}. It is used in trace-stitching to
restore the original control flow.

\subsubsection{Function call.}

To reduce the compilation code size, our baseline \jit{} compiler does not inline a
function call.

\paragraph*{When tracing \texttt{CALL} instruction at \circled{4},} it does not
follow the destination of \texttt{CALL} but emits only a call instruction since
subroutines are decorated with \texttt{dont\_look\_inside}.

\subsubsection{Function return.}

\paragraph*{When tracing \texttt{RET} at \circled{5},} first, the baseline tracer checks
whether \texttt{traverse\_stack} is empty or not. If not empty, it restores a saved program
counter and continues to trace. Otherwise, it executes \texttt{RET} instruction. The
implementation is shown in Listing~\ref{lst:def_ret}, and the behavior is almost same
to \texttt{JUMP}.

We finally get the following trace as shown in
Listing~\ref{lst:trace_from_method_traversal_interp}. Note that it is still linear, so we
will cut and stitch the generated trace to restore the original control flow.

\begin{lstlisting}[language=Python,label={lst:trace_from_method_traversal_interp},
caption={The temporarily generated trace from a method-traversal interpreter.},float=t]
[p0]
i1 = call_i(ConstClass(tla_DUP, p0))
i2 = call_i(ConstClass(tla_CONST_INT, p0, 1))
i3 = call_i(ConstClass(tla_GT, p0, 2))
i4 = call_i(ConstClass(_is_true, p0, 4))
guard_true(i4) [p0]
i5 = call_i(ConstClass(tla_CONST_INT, p0, 7))
i6 = call_i(ConstClass(tla_SUB, p0))
i7 = call_i(ConstClass(cut_here, 8))
i8 = call_i(ConstClass(tla_CALL, p0, 10))
i9 = call_i(ConstClass(tla_RET, p0, i8))
leave_portal_frame(0)
finish(i9)
\end{lstlisting}

\subsection{Trace Stitching}\label{sec:trace_stitching}

The obtained trace by tracing method-traversal interpreter is a linear execution
path, since the tracer is led to track all paths by the interpreter.
For correct execution, we propose trace stitching,
which is a technique to reconstruct the original control flow.

Figure~\ref{fig:overview_trace_stitching} shows how trace stitching works, and \fbox{1} --
\fbox{5} indicate its working flow.

\begin{description}
\item [\fbox{1}: the tailor cuts where \texttt{cut\_here} indicates] to handle each branch
  as a separate trace. In Figure~\ref{fig:overview_trace_stitching}, the tailor cuts the
  node B in Figure~\ref{fig:overview_trace_stitching} that \texttt{cut\_here} points to;
\item [\fbox{2}: the tailor restores the conditional branch] by compiling the trace E
  -- F as a bridge. When compiling as a bridge, the tailor emits a label L and rewrites
  the definition of an original guard failure that is placed at B;
\item [\fbox{3}: the tailor restores \texttt{JUMP} instruction] at the bottom of D. After
  that,
\item [\fbox{4}: it copies variables and instructions] that are not in the scope of the branch B --
  E -- F for run-time correctness. Finally,
\item [\fbox{5}: the tailor folds or removes] constants or unused
  variables/instructions, respectively.
\end{description}


As a result, we get the trace tree as shown in the rightest side of
Figure~\ref{fig:overview_trace_stitching}. Inside the \rpython{}, the trace tree is
represented as two traces shown in Listing~\ref{lst:traces_after_tailoring}. There
is no linear trace, but one trace and bridge are connected with a guard failure.
If \texttt{guard\_true(i4)} is failed, the control goes to the Bridge 1 and executes it.

\begin{lstlisting}[language=Python,float=t,label={lst:traces_after_tailoring},
caption={Tailored traces. One linear trace is converted into one trace and one bridge,
and they are connected with a guard failure.}]
# Loop 1, token number is 13458300
[p0]
i1 = call_i(ConstClass(tla_DUP, p0))
i2 = call_i(ConstClass(tla_CONST_INT, p0, 1))
i3 = call_i(ConstClass(tla_GT, p0, 2))
i4 = call_i(ConstClass(_is_true, p0, 4))
guard_true(i4) [p0] # pointing to Bridge 1
i5 = call_i(ConstClass(tla_CONST_INT, p0, 7))
i6 = call_i(ConstClass(tla_SUB, p0))
# targeting to its own top
jump(p0, descr=TargetToken(13458300))

# Bridge 1, token number is 1345340
[p0]
i8 = call_i(ConstClass(tla_CALL, p0, 10))
i9 = call_i(ConstClass(tla_RET, p0, i8))
leave_portal_frame(0)
finish(i9)
\end{lstlisting}

\begin{figure*}[t!]
  \centering
  \begin{subfigure}{.42\linewidth}
    \centering
    \includegraphics[width=\linewidth]{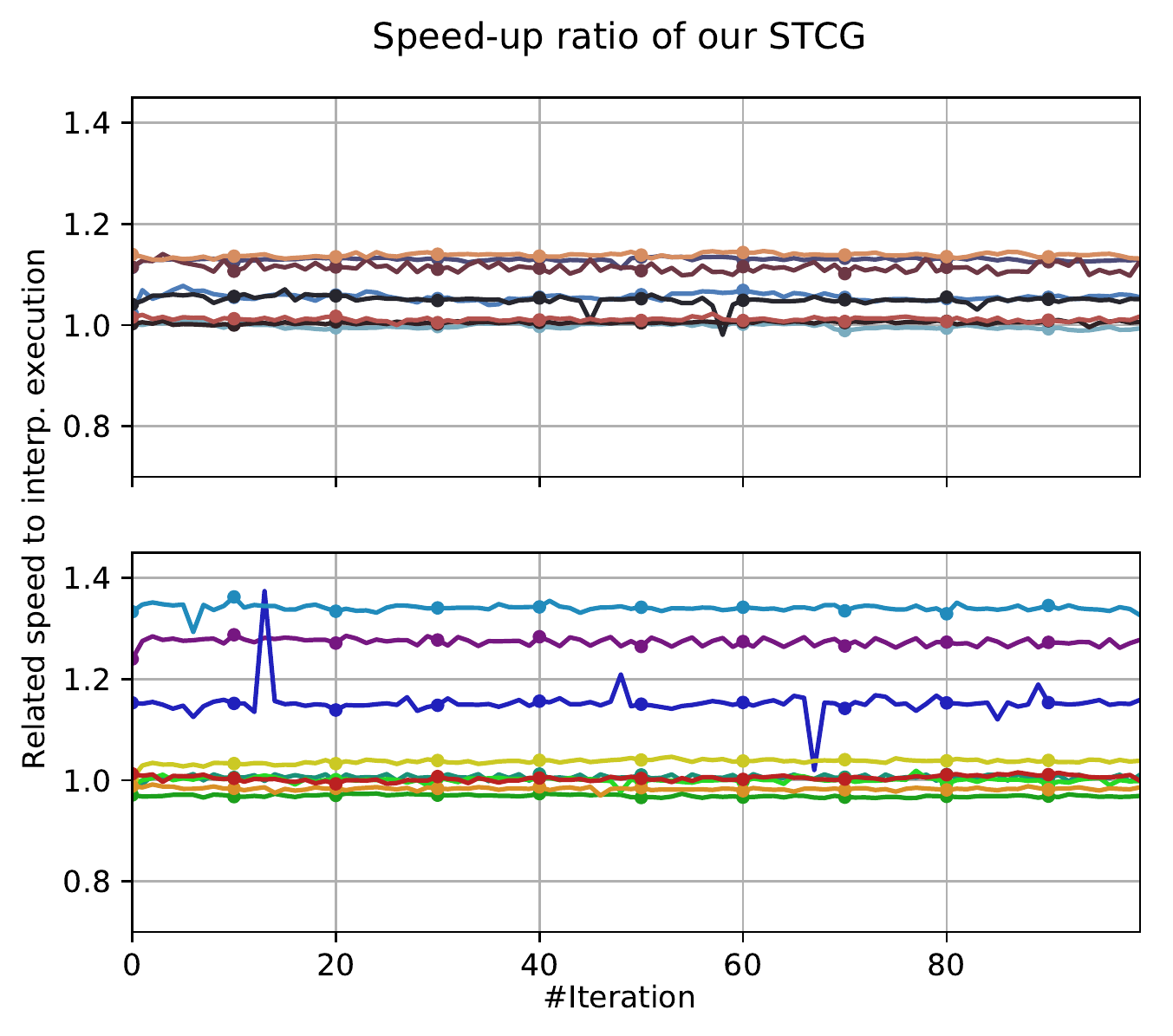}
  \end{subfigure}
  \begin{subfigure}{.525\linewidth}
    \centering
    \includegraphics[width=\linewidth]{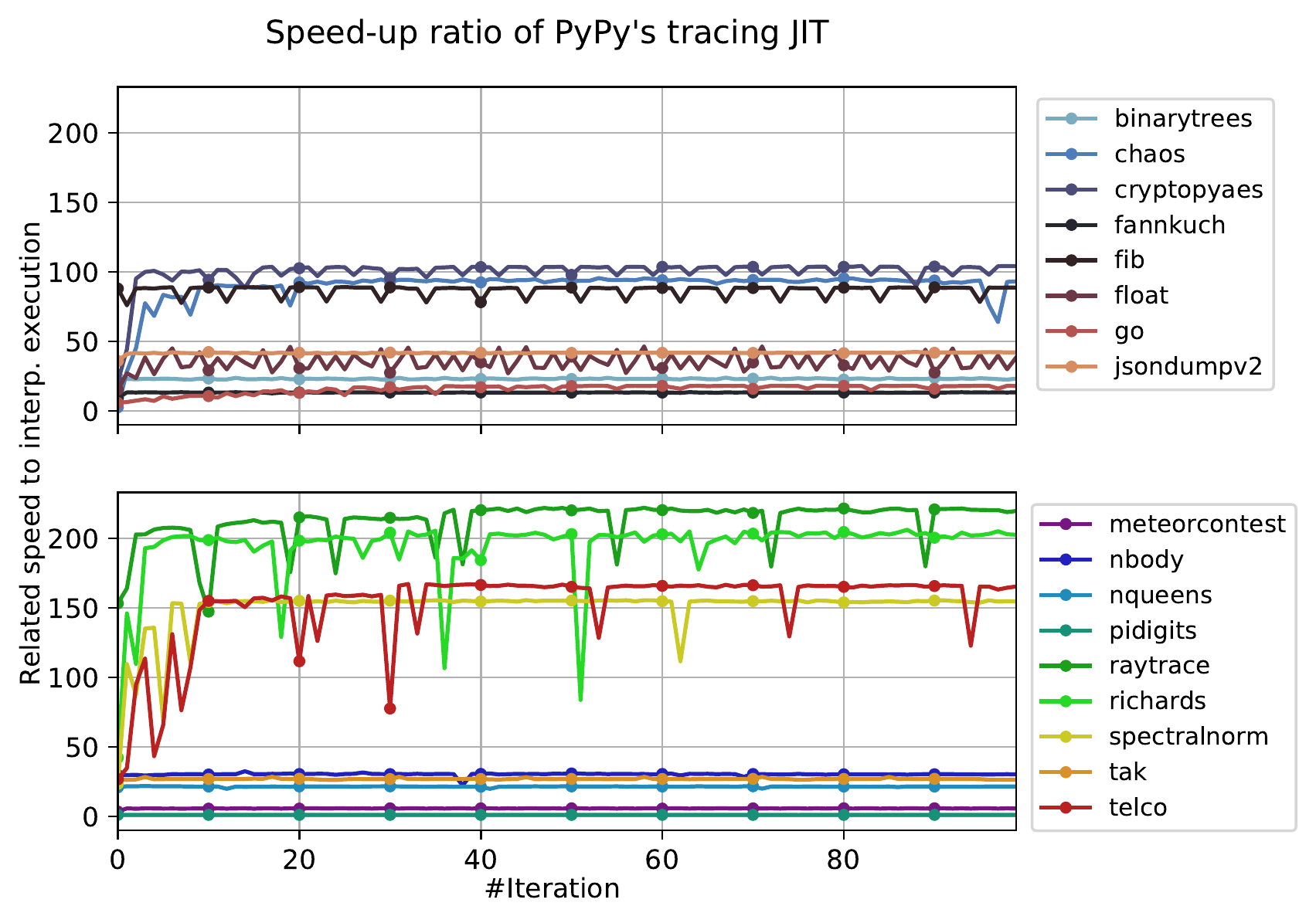}
  \end{subfigure}
  \caption{
    Speed-up ratio of STCG (left-hand side) and PyPy's tracing JIT compiler (right-hand
    side) related to the interpreter. They are executed on PyPy's original micro benchmark
    suite plus our original ones. X-axis and Y-axis mean every iteration and speed-up
    ratio standardized to interp. execution, respectively. Dots are plotted every five
    iterations.}
  \label{fig:threaded-code-result-per-iter}
\end{figure*}

\section{Performance of Simulated Threaded Code Generation}
\label{sec:discussion}

In this section, we experimentally evaluate the potential performance of our threaded code
generation by simulating the behavior with PyPy. We here compare the threaded code
performance against the interpreter performance, although we are also interested in that
against the best possible threaded code performance. It would be an interesting future work
to implement and compare different threaded code generations in real-world languages like
Python.


In Section~\ref{sec:compilation-tactic-of-baseline-jit}, we described the idea of threaded
code generation that enables a baseline compilation with a meta-tracing \jit{}
compiler. The question arises whether the technique is effective at runtime or not. To
answer the question, we measured \jit{} compilation time and code size of traces of
PyPy's tracing \jit{} compiler and our simulated threaded code generation (SMTG),
respectively. In addition, we compared the potential performance of the two following
executions: PyPy 3.7 with SMTG and interpreter-only execution.

\subsection{Simulated Threaded Code Generation (STCG) in PyPy}\label{sec:simulate-threaded-code-gen}

To measure the potential performance of our threaded code generation, we need to
reproduce its behavior on PyPy. The brief ideas of the \emph{simulated threaded code
generation (STCG)} are;

\begin{description}
\item[Idea 1.] All subroutines are not inlined, but call instructions to subroutines are
  left.
\item[Idea 2.] Tracing all paths of a target program area \emph{at once}.
\end{description}

Idea 1 can be easily reproduced by adding \texttt{dont\_look\_inside} to the PyPy
interpreter manually. The problem is how to reproduce idea 2. The current PyPy doesn't
have such a function, but it has a guard failure. A guard failure is a runtime check to
ensure the correctness of the generated trace. When the number of failing a guard
surpasses a threshold, a tracing JIT starts to trace the destination of a guard and
connects the original trace and the generated trace from a guard. Then, if we run
programs with enough time, all runtime paths are eventually traced by a guard failure. Therefore, we
can reproduce the behavior of idea 2 by running the benchmarks for a long time.

\subsection{Setup}

In this section, we explain the environment and how we performed our preliminary experiments.

\subsubsection{System}

We conducted the preliminary benchmark on the following environment; CPU: Ryzen 9
5950X, Mem: 32GB DDR4-3200MHz, OS: Ubuntu 20.04.3 LTS with a 64-bit Linux kernel
5.11.0-34-generic.

\subsubsection{Implementation}

We used the original PyPy 3.7 versioned
7.3.5\footnote{\url{https://downloads.python.org/pypy/pypy3.7-v7.3.5-linux64.tar.bz2}},
and our modified PyPy 3.7 with
STCG\footnote{\url{https://foss.heptapod.net/pypy/pypy/-/tree/branch/py3.7-hack-measure-bytecode-dispatch}}.

\subsubsection{Programs for Experiments}

You can find all benchmark programs here\footnote{
  \url{https://foss.heptapod.net/pypy/benchmarks/-/tree/topic/python3_benchmarks/bitbucket-pr-5}}.
We chose all benchmarks that can be executed without any other libraries. Especially,
\texttt{fib} and \texttt{tak} are programs causing the path-divergence problem.

\subsubsection{Methodology}

We conducted two experiments on PyPy's original micro benchmark suite plus our original
ones;

\begin{description}
\item[Experiment 1.] Measuring the overhead of tracing and compilation in our STCG.
\item[Experiment 2.] Measuring the stable speeds of our STCG.
\end{description}

\begin{figure*}[t!]
  \centering
  \includegraphics[width=.95\linewidth]{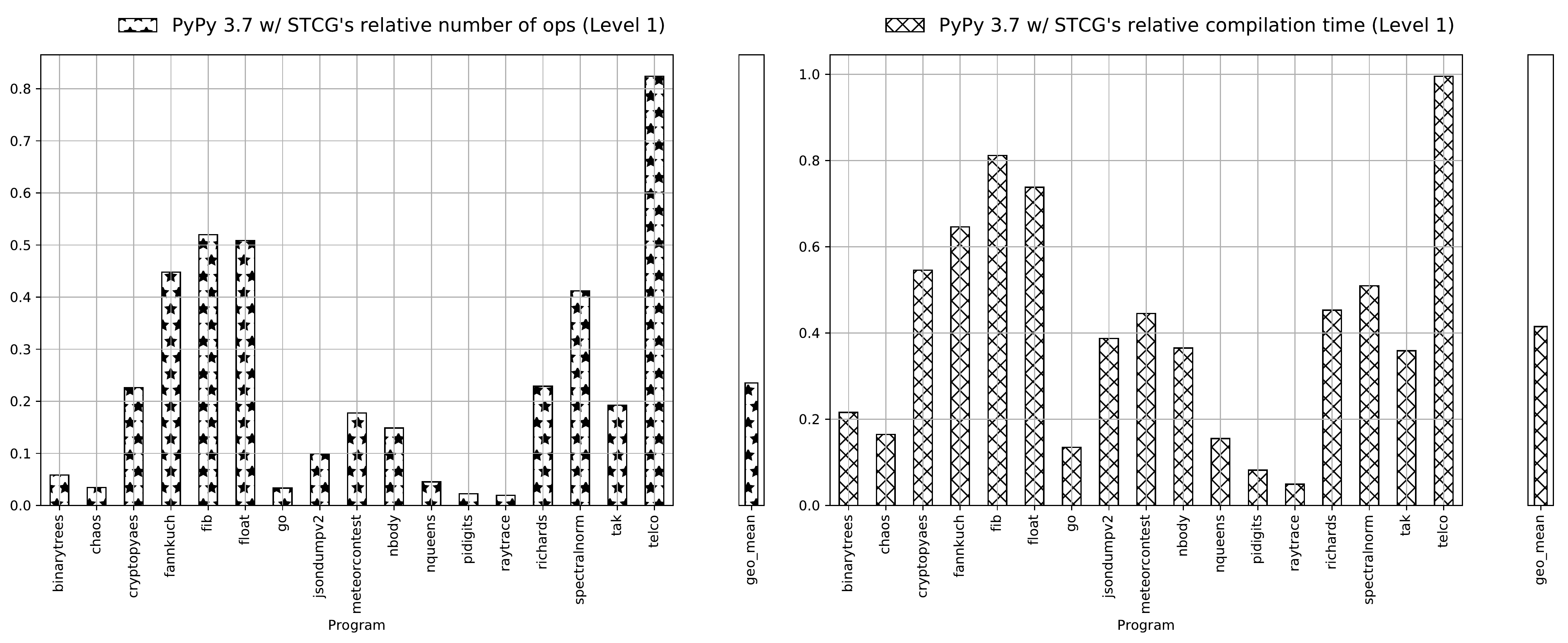}
  \caption{The results of the size of traces to compile and compilation time including
    tracing. In all results the Y-axis means PyPy 3.7 with our simulated threaded code
    generation (STCG)'s relative value to PyPy 3.7--7.3.5's tracing JIT compiler. The
    X-axis stands for the name of every program. The left-hand side shows the relative
    trace size, and the right-hand size is the relative compilation time. Lower is
    better.}
  \label{fig:trace-size-and-comp-time}
\end{figure*}

\begin{figure*}[t!]
  \centering
  \includegraphics[width=.95\linewidth]{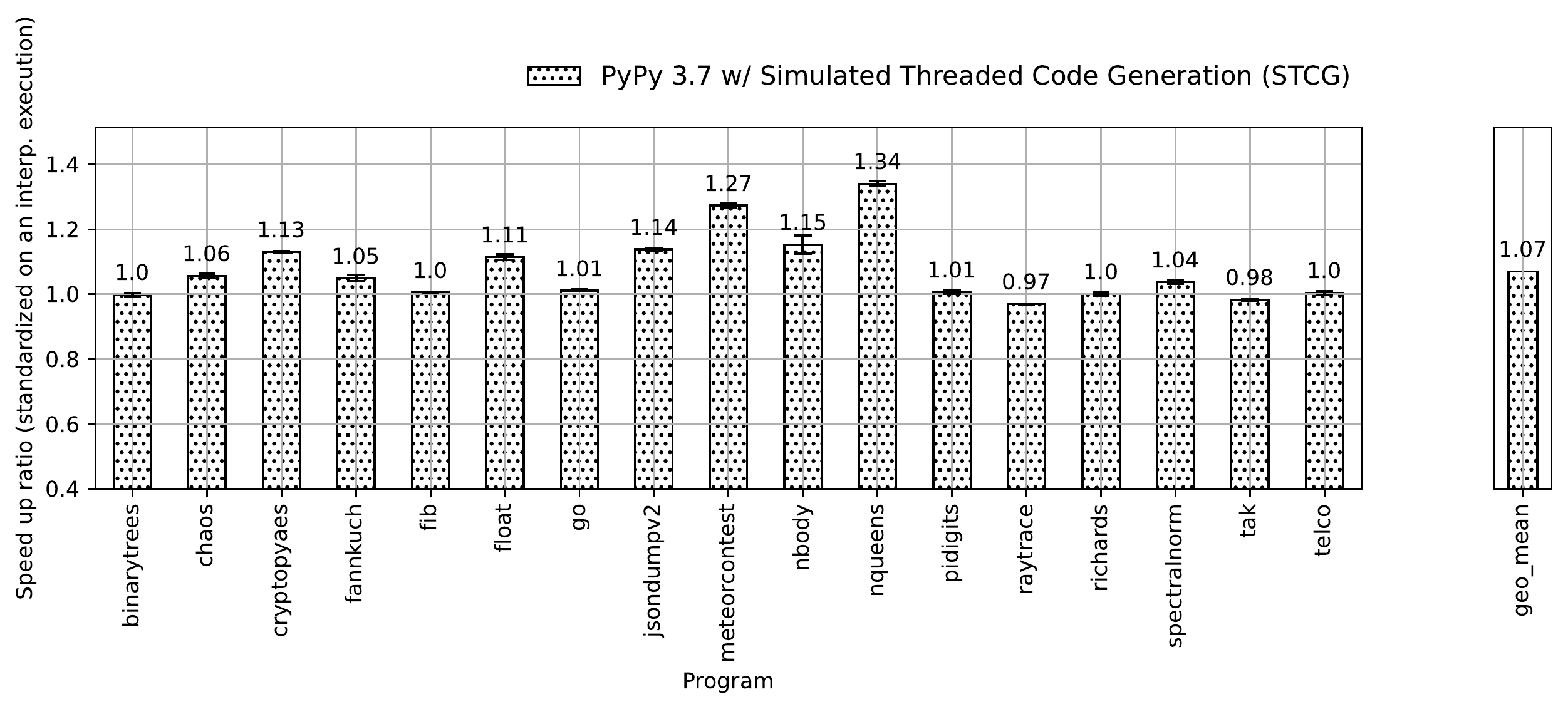}
  \caption{The results of a preliminary benchmark experiment. In all results the Y-axis
    means speed up ratio of the threaded code generation comparing to the interpreter-only
    execution, and the X-axis stands for the name of every program. The error bars mean
    standard deviations. Higher is better.}
  \label{fig:threaded-code-result}
\end{figure*}

\paragraph{Experiment 1.}

To measure the overhead of tracing and compilation, we used PyPy 3.7--7.3.5 with a tracing
JIT and our STCG. We measured their compilation time and the size of traces to compile,
and normalized them to PyPy with a tracing JIT. The compilation time includes
tracing. Note that the implementation of our full-fledged threaded code generation on PyPy
is ongoing, so we simulate the behavior (we describe how to do it in Section~\ref{sec:simulate-threaded-code-gen}).
The results are shown in Figure~\ref{fig:trace-size-and-comp-time}.

\paragraph{Experiment 2.}

To compare a stable speed, we compared the STCG on PyPy 3.7 with the interpreter-only
execution. The interpreter-only execution means that we turn off the JIT compilation by
passing \verb|--jit off| when running scripts. We calculated the averages and
standard deviations of the STCG normalized to the interpreter-only execution.
The results are shown in Figure~\ref{fig:threaded-code-result}.

We set the max iteration count 100 from the results that plot the related speed-up ratio
in the STCG and PyPy's tracing \jit{} as shown in
Figure~\ref{fig:threaded-code-result-per-iter}. From those results in the STCG, we can
confirm that almost all programs except reach their stable state after the 5th iteration.
In addition, the PyPy's tracing \jit{} reaches its stable speed after the 30th iteration.
Experiment 1 requires a number of operations and times for compilation, and
experiment 2 needs STCG's stable speed; in this context, we decide that the max iteration
count 100 is enough to reach the stable speed. Thus, in experiment 2, we exclude the first
5 iterations for calculating the average value of every program's stable speed.

\subsection{Result of Experiment 1: The Overhead of Our STCG}
\label{sec:result-exp-1}

The objective of this experiment is to potentially evaluate the start-up time of our
simulated threaded code generation. The results are shown in
Figure~\ref{fig:trace-size-and-comp-time}. On average, in the case of trace sizes
to compile, PyPy 3.7 with STCG is about 78 \% smaller than PyPy 3.7--7.3.5 with a tracing
\jit{}, and 13 of 17 programs are about 50 \% smaller than PyPy's tracing \jit{}.
In addition, in the case of compilation time, PyPy 3.7 with STCG is about 60 \% shorter
than PyPy 3.7--7.3.5 with a tracing \jit{}. 13 of 17 programs, that are same to the case of
the size of traces to compile, are 60 \% shorter than PyPy's tracing \jit{}.
However, PyPy 3.7 with STCG size of traces and compilation time on \texttt{nbody} is almost
the same as that of PyPy's tracing \jit{}. This program computes the N-body simulation
with a matrix calculation. This calculation is implemented as a big for-loop, so there is
less effect on performing threaded code generation than full-optimized tracing.

\subsection{Result of Experiment 2: The Stable Speed}
\label{sec:result-exp-2}

The results are summarized in Figure~\ref{fig:threaded-code-result} (their values in every
iteration are shown in Figure~\ref{fig:threaded-code-result-per-iter}). The results of PyPy
3.7 simulated threaded code generation are normalized to the interpreter only
execution. On average, STCG is 7\% faster than the interpreter only. PyPy 3.7 with STCG is
over 4 \% faster in 9 of the 17 benchmarks, and $\pm$ 3 \%  faster in 8 of the 17
benchmarks. In particular, \texttt{meteorcontest} and \texttt{nqueens} are from about 27\%
to 34 \% faster than the interpreter.

\subsection{Discussion}

In experiment 1, there is a relation between the size of traces and the
compilation time. Our simulated threaded code generation can reduce the size of traces and
compilation time, so we can use it for reducing the start-up time.

Moreover, programs with the path-divergence problem (\texttt{fib} and \texttt{tak}) are at
least 96 \% smaller and 80 \% faster in trace size and compilation time, respectively. In
general, when the path-divergence problem occurs, retracing often happens, and too many
traces overlap each other and lead to high overhead in run-time performance. However, the
result shows that the STCG traces and compiles only a primary hot function, so the trace
sizes and compilation time are much smaller and shorter than a tracing JIT. Thus, we can
say that a method-based threaded code can reduce the trace size and compilation time.

From both experiments, we can infer that our method-based threaded code generation will
bring some benefits to a start-up performance. To make the technique more effective, we
should select functions that have similar structure to \texttt{meteorcontes} and
\texttt{nqueens} as well as programs with the path-divergence problem. In those programs,
much part of one primary solver function with complex conditional branches is executed
inside the main loop, but the other functions are not. In other words,
during solving conditions, instead of running a main single region over and over, some
regions are sometimes run randomly. This execution model
potentially causes the path-divergence problem. Thus, a method-based threaded code
generation can work effectively on such programs. To enhance the
effectiveness of our method-based threaded code generation with this assumption, we need
to select programs with complex conditional branches inside a long iteration in addition
to programs which indeed cause the path-divergence problem.

\paragraph{Limitation of Threaded Code Generation.}

Threaded code generation is placed at the initial compilation tier, so the compilation
limits further optimizations. For example, since the compilation does not inline an
instruction handler but leaves a call instruction to that. We notice that there are gaps
between this baseline \jit{} compilation and the tracing \jit{} that \rpython{}
provides. Thus, this gap suggests that we need several optimizations between the baseline
and tracing \jit{}s.

To allow further optimizations like allocation removal, we are going to implement higher
levels of baseline \jit{} compilation. For instance, the tier-2 baseline \jit{} just
inlines a stack manipulation, but other operations are not. The tier-3 baseline \jit{}
inlines auxiliary methods but others are not inlined. We are able to realize those levels
at low cost by placing \texttt{dont\_look\_inside} to each method header.

\section{Related Work}
\label{sec:relatedwork}

The trade-off between compilation time and peak performance has been actively
discussed in the context of compiler implementation. For long-running applications such as
server-side programs, we would accept a long compilation time. In contrast, for short-term
applications such as GUI programs or batch processing programs, we would require a better
response time, hence we usually apply a baseline \jit{} compiler at first.

The Java \hotspot{} VM has two \jit{} compilers: the server
compiler~\cite{paleczny2001java} and the client
compiler~\cite{Kotzmann:2008:10.1145/1369396.1370017}. The server compiler is a
highly optimizing compiler and is tuned to gain a faster peak-time performance with
lower compilation speed. On the other hand, the client compiler is a \jit{}
compiler designed for low start-up time and small memory footprint.

The Firefox baseline compiler~\cite{firefoxbaselinejit} is a warm-up compiler used in
the IonMonkey JavaScript \jit{} compiler~\cite{ionmonkey}. Firefox's baseline \jit{} is
designed to work as an intermediate layer between interpretation and highly optimizing
\jit{} compilation. Firefox used different \jit{} compilers, JaegerMonkey and
IonMonkey, depending on the situation, but it had several issues. For
example, the calling conventions of the two compilers are different. Moreover,
JaegerMonkey itself has a too complex structure to easily extend. Firefox's
baseline \jit{} compiler is designed to solve these issues. Its baseline \jit{} compiler
is simpler than other compilers, but runs 10--100 times faster than
interpretation.

The Liftoff~\cite{liftoff} is a baseline \jit{} compiler for V8 and WebAssembly. V8 has
an older \jit{} compiler called TurboFan, but its compilation process is complicated,
and it consumes longer compilation time. Liftoff makes the code quality secondary in
order to achieve a faster start-up time, which is the key difference from the TurboFan
compiler.

The Safari's JavaScript engine, JavaScriptCore, has 4-tier optimization levels in its
\vm{}~\cite{javascriptcore}. The engine consists of low-level interpreter
(tier-1), baseline \jit{} (tier-2), data-flow graph \jit{} (DFG, tier-3), and forth-tier
\jit{} (FTJ, tier-4) compilers. Execution firstly enters the interpreter-tier, and
level-shifting between every \jit{} compiler can be executed by on-stack replacement
(OSR)~\cite{Holzle1994:10.1145/191080.191116}. In particular, the baseline \jit{} does not
apply serious optimizations but just eliminates the interpretation overhead. Polymorphic
inline caching (PIC)~\cite{Holze1991:10.5555/646149.679193} is used in the baseline \jit{}
a classic optimization technique to remove dynamic method dispatching, and profiling
information gathered when PIC is performed are passed to higher-level \jit{} compilers.

\section{Conclusion and Future Work}
\label{sec:conclusion}

\subsection{Conclusion}

In this paper, we proposed the idea of a method- and threaded-code-based \rpython{}'s
baseline \jit{} compiler and how to implement them on top of RPython.
The essential technique is trace tailoring that consists of the method-traversal
interpreter and trace stitching. A method-traversal interpreter is an interpreter
design that tricks the trace to follow all paths of a target function. Trace
stitching rebuilds a trace tree from a resulting trace generated from a
method-traversal interpreter, aiming to restore the original control flow. In average, our
experiments report that threaded code can reduce the size of traces and compilation time
by about 80 \% and 60 \%, respectively. It can run 7 \% faster than the interpreter-only
execution in the case of a stable speed.

\subsection{Future Work}

\paragraph{Multitier Adaptive Compilation.}

In the viewpoint of code quality and compilation time, our threaded code generation is
placed at an interpreter execution and tracing JIT compilation. We would
connect them and shift the compilation level depending on a target program. For example,
we start baseline \jit{} compilation before applying tracing \jit{} compilation. Then, when
we find a program fragment that is suitable for tracing \jit{} compilation, we use a
tracing \jit{} compiler instead of a baseline \jit{} compiler. In the future, we would
realize such an adaptive compilation strategy on \rpython{}.

\paragraph{Implementing Threaded Code Generation on PyPy.}

Currently, we designed a method-traversal interpreter for a tiny language and
created a compiler that could emit a trace tree that contains only call instructions
to subroutines. Our next task is to implement our idea on PyPy.  By comparing with the
original tracing \jit{} in \rpython{} and PyPy, we will see how much start-up time and
memory footprint can be reduced in practice. Finally, we will verify the effectiveness of
our baseline \jit{} on production-level applications by using the PyPy that has a baseline,
method (that will be extended from baseline \jit{}), and tracing compilation
strategies. Given this context, we will implement it in Python with the PyPy interpreter
to run production-level benchmarks.

\bibliography{main}

\section*{About the authors}
\shortbio{Yusuke Izawa}{is a Ph.D. student at the Tokyo Institute of Technology
  (Japan). \authorcontact[https://www.yuiza.org]{izawa@prg.is.titech.ac.jp}}

\shortbio{Hidehiko Masuhara}{is a professor at the Tokyo Institute of Technology.
  \authorcontact[]{masuhara@acm.org}}

\shortbio{Carl Friedrich Bolz-Tereick}{is PyPy/RPython contributor
  and scientific employee at Heinrich-Heine-Universität
  Düsseldorf. \authorcontact[https://cfbolz.de]{cfbolz@gmx.net}}

\shortbio{Youyou Cong}{is an assistant professor at the Tokyo Institute of Technology.
  \authorcontact[]{cong@c.titech.ac.jp}}
\end{document}